\documentstyle[aps,epsf,prl,epsfig,twocolumn]{revtex}

\begin{document}

\title{\bf Kibble-Zurek Mechanism in the Ginzburg Regime: \\
             Numerical Experiment in the Ising Model       }

\author{ Jacek Dziarmaga }

\address{ Intytut Fizyki Uniwersytetu Jagiello\'nskiego,
              ul.~Reymonta 4, 30-059 Krak\'ow, Poland           }

\date{ October 22, 2003 }

\maketitle

\begin{abstract}
Kibble-Zurek mechanism is a theory of defect formation in a non-equilibrium 
continuous phase transition. So far the theory has been successfully tested
by numerical simulations and condensed matter experiments in a number of systems
with small thermal fluctuations. This paper reports first numerical test
of the mechanism in a system with large thermal fluctuations and strongly 
non-mean-field behavior: the two dimensional Ising model. The theory predicts
correctly the initial density of defects that survive a quench from the 
disordered phase. However, before the system leaves the Ginzburg regime of large 
fluctuations most of these defects are annihilated and the final density is 
determined by the dynamics of the annihilation process only. 
\end{abstract}



{\bf Introduction.---} In a system with a continuous phase transition an
adiabatic change of a parameter of the system, like e.g. temperature,
pressure or a coupling constant in a Hamiltonian, can drive the system
from a disordered phase to an ordered one. A classic example is the
paramagnet-ferromagnet transition in the two dimensional (2D) Ising model.
Thermodynamics of continuous phase transitions has been intensively
explored over many years. Two mayor achievements: the solution of the
Ising model by Onsager and the renormalization group of Wilson were
rewarded with a Nobel Prize in physics. The RG formalism revealed deep
connections between statistical mechanics and quantum field theory.

A candidate theory of non-equilibrium phase transitions is the
Kibble-Zurek mechanism (KZM) \cite{Kibble,Zurek}. Kibble pointed out
\cite{Kibble} that in a non-equilibrium transition there is no time to
fully develop the long range order characteristic for the ordered phase.
As a result, the final state of the system is a mosaic of finite size
ordered domains with different orientations of the order parameter in
every domain. In a topologically non-trivial case this disorder takes the
form of a finite density of topological defects. This qualitative idea was
quantified more by Zurek \cite{Zurek}. Zurek mechanism is a combination of
two basic facts: (1) a divergent correlation length
\begin{equation}
\xi~\approx~\xi_0~|\epsilon|^{-\nu}~,
\label{nu}
\end{equation}
where $\epsilon$ is a dimensionless distance from the critical point,
$\nu$ is a critical exponent, and $\xi_0$ is a microscopic length scale, 
and (2) the critical slowing down or divergent relaxation time
\begin{equation}
\tau~\approx~\tau_0~|\epsilon|^{-y}~.
\label{y}
\end{equation}
Here $\tau_0$ is a microscopic time scale. Because of the divergent relaxation 
time any finite rate transition is a non-adiabatic phase transition: 
sufficiently close to the critical point (where $\epsilon=0$) the system is 
too slow to react to the changing external parameter $\epsilon(t)$. Close to 
$\epsilon=0$ we can linearize 
\begin{equation}
\epsilon(t)~=~\frac{t}{\tau_Q}~. 
\end{equation}
The relaxation time (\ref{y}) equals the transition rate 
$|\epsilon/\frac{d\epsilon}{dt}|$ at 
$\epsilon_Z\approx(\tau_0/\tau_Q)^{\frac{1}{y+1}}$ when the correlation 
length (\ref{nu}) is
\begin{equation}
\xi_Z~\approx~
\xi_0~\left( \frac{\tau_Q}{\tau_0} \right)^{\frac{\nu}{y+1}}~.
\label{hatxi}
\end{equation}  
This Zurek length is the average size of the ordered domains after the
phase transition and it determines the initial density of topological
defects frozen into the ordered phase after a non-adiabatic continuous
phase transition.

The original motivation for Kibble and Zurek were symmetry breaking phase 
transitions in cosmology. The random topological defects arising in such 
transitions might provide initial seeds for structure formation in the early
Universe \cite{book}. However, the universality of phase transitions makes these 
ideas also relevant for a wide variety of condensed matter systems where they 
can be verified by experiment. 

The KZM prediction (\ref{hatxi}) is supported by a number of numerical
simulations \cite{num}. However, as a result of finite numerical resources
these numerical data are limited to fast quenches (small $\tau_Q$) with a
large $\epsilon_Z$ in the regime of small fluctuations where one can use
the mean field (MF) value of the critical exponent $\nu_{\rm MF}=\frac12$.  
KZM is also supported by experiments in systems with small fluctuations
like superfluid helium 3 \cite{he3}, low $T_c$ superconductors
\cite{monaco}, and fast quenches in high $T_c$ superconductors
\cite{fasthighTc}. In contrast, experiments in systems with large
fluctuations like helium 4 \cite{he4} or slow quenches in high $T_c$
superconductors \cite{slowhighTc} are inconclusive. Rivers suggested
\cite{Rivers} that vortices created in the helium 4 experiment \cite{he4}
disappear in a faster than expected annihilation. Due to technical
difficulties the analytic calculations in Ref.\cite{Rivers} eventually
resort to a linearization equivalent to the mean-field theory. It is
suggested there that beyond this linearized theory close to the critical
point the annihilation rate is divergent. However, simulations in
Ref.\cite{Bett} show that this effect may be not as dramatic as
anticipated in Ref.\cite{Rivers}.  These authors suggest that because of
the critical slowing down the annihilation rate close to $\epsilon=0$ may
in fact vanish. Due to limited numerical resources the numerical evidence
in Ref.\cite{Bett} is rather indirect. To summarize, the problem of KZM in
the Ginzburg regime of large fluctuations has been recognized
\cite{Rivers} but is far from being settled.

At the moment we do not have any condensed matter or numerical experiment
supporting KZM for large fluctuations and at the same time this is the
regime where KZM in principle can be questioned on general 
grounds. The
argument leading to Eq.(\ref{hatxi}) implicitly assumes that close to the
critical point the divergent correlation length $\xi$ in Eq.(\ref{nu}) is
the only relevant length scale.  However, as is well known
\cite{Goldenfeld} but not quite generally appreciated, if $\xi$ were the
only length scale, then, on dimensional grounds, all critical exponents
would take their mean field values. As they do not (for example, in the 2D
Ising model $\nu=1$ instead of the mean field $\nu_{\rm MF}=\frac12$),
then both $\xi$ and the microscopic $\xi_0$ must be relevant. With two
relevant length scales the dimensional argument alone is not sufficient to
determine the initial density of defects.

In this paper I report first numerical test of KZM for large fluctuations.
As the critical regime is numerically demanding (large $\xi$ means large
lattice and large $\tau$ means long time) I chose the simplest possible
model - the celebrated 2D Ising model. This simple model has $\nu=1$
clearly different from the mean field $\nu_{\rm MF}=1/2$, and it has no
regime where the MF theory might be at least remotely accurate. It is a
perfect testing ground for KZM.


{\bf Ising model with Glauber dynamics.---} Hamiltonian of the ferromagnetic 
Ising model in 2D is
\begin{equation}
H~=~-~\sum_{\langle i,j \rangle}~S_i~S_j~.
\label{HIsing}
\end{equation}
Spins $S_i\in \{ -1,+1 \}$ sit on a 2D $N\times N$ lattice with periodic
boundary conditions, $\langle i,j \rangle$ means a pair of
nearest neighbor sites. The microscopic lengthscale $\xi_0=1$ is the
lattice spacing. In all the following numerical simulations a 
$1024\times 1024$ lattice was used.

To study non-equilibrium phase transitions the Ising model has to be 
supplemented with dynamics. The standard choice is Glauber dynamics also known 
as Monte-Carlo with a heat bath \cite{neural}. In the Glauber algorithm every 
time step consists of the following sub-steps:
\begin{itemize}
\item choose a random spin $S_i$ from the lattice,
\item calculate its local field $h_i=-\sum_{j~{\rm n.n.}~i}S_j$,
\item calculate a probability $P=\exp(\beta h_i)$,
\item choose a random number $r\in [0,1]$,
\item if $r>P$ then set $S_i=+1$, else set $S_i=-1$.
\end{itemize}
Here $\beta$ is an inverse temperature of the heat bath. This algorithm relaxes 
the state of the Ising model towards thermal equilibrium at a given $\beta$ 
\cite{neural}. On average it takes $N^2$ steps to update the state of $N^2$ 
spins on the lattice. These $N^2$ steps define the microscopic time scale 
$\tau_0$ which I set equal to $1$.

The Ising model with Glauber dynamics belongs to the same universality
class as the $\phi^4$ model with noise $\eta$
\begin{equation}
\tau_0\frac{\partial}{\partial t}\phi=
\xi_0^2\nabla^2\phi-\lambda(\phi^2-1)\phi+\eta
\label{phi4}
\end{equation}
so often employed in the numerical simulations of KZM \cite{num}. Here the
continuum real field $\phi$ is a coarse grained lattice spin $S_i$. The
Ising model is an efficient ``molecular dynamics'' version of the $\phi^4$
field theory (\ref{phi4}).

\begin{figure}
\epsfig{file=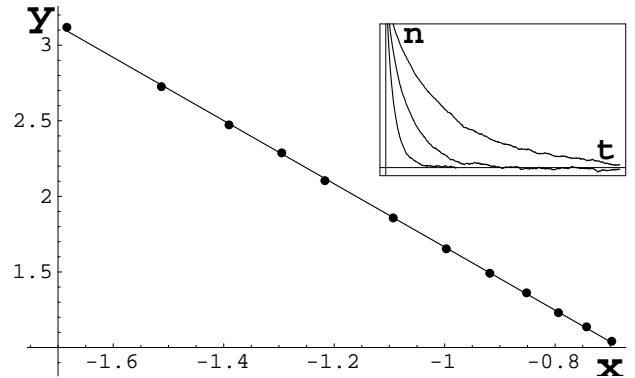, width=8.6cm, clip=}
\vspace*{0.1cm}
\caption{$y=\log_{10}\tau$ as a function of $x=\log_{10}(\beta_c-\beta)$.  
The $\tau$s are the best fits to the exponential decays of magnetization
shown in the insert. The solid line is the best linear
fit with a slope of $y=2.09\pm 0.02$. }
\label{M}
\end{figure}

{\bf Relaxation time.---} In order to estimate the exponent $y$ in
Eq.(\ref{y})  the relaxation time $\tau$ was measured for several values
of $\beta<\beta_c$.  For each $\beta$ the Ising model was initially
prepared in a fully polarized state with all $S_i=1$, and then its average
magnetization $M=\sum_iS_i/N^2$ relaxed towards the equilibrium at $M=0$,
see the insert in Fig.\ref{M}.  Each magnetization decay was fitted with
an exponent $M=\exp(-t/\tau)$. The best fits of $\tau$ are shown in the
double logarithmic Fig.\ref{M} as a function of $\beta_c-\beta$. The slope
of the linear fit in Fig.\ref{M} gives an estimate of $y=2.09\pm 0.02$.

\begin{figure}
\epsfig{file=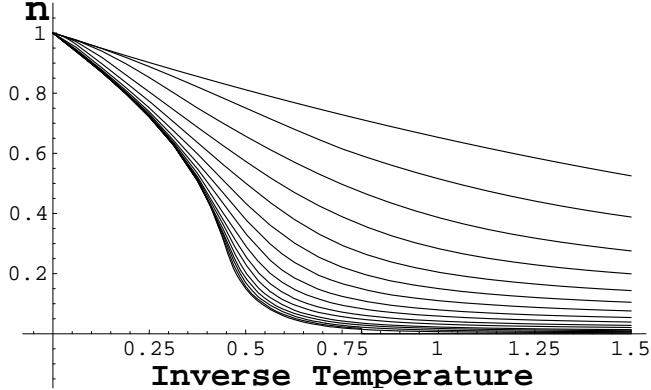, width=8.6cm, clip=}
\vspace*{0.1cm}
\caption{ Total density $n(\beta)$ of domain walls separating positive and
negative $S_i$ as a function of $\beta$ for several values of the
quench time $\tau_Q=2,4,8,\dots,65536$ (from top to bottom). For the 
initial state
of random spins the density is normalized to $n=1$. For large $\tau_Q$ the
plots tend to the equilibrium density of defects $n_{eq}(\beta)$ which is
finite even for $\beta>\beta_c=0.4407$ in the Ginzburg regime of large
fluctuations.}
\label{n}
\end{figure}

{\bf Quenches.---} Phase transitions were simulated with a linear ramp of the 
inverse temperature
\begin{equation}
\beta(t)~=~1.5~\frac{t}{\tau_Q}~.
\label{ramp}
\end{equation} 
The initial state at $t=0$ was a state with random mutually uncorrelated spins -
the state of equilibrium at $\beta=0$. Fig.\ref{n} shows density of domain walls 
separating positive $S_i$ from negative $S_i$ as a function of $\beta$ for a 
number of different transition times $\tau_Q$. The 
critical point is $\beta_c=0.4407$. For large $\tau_Q$ the density plots 
approach the 
equilibrium density $n_{eq}(\beta)$. Note that the equilibrium density 
$n_{eq}(\beta)$ of domain walls remains nonzero even for $\beta>\beta_c$. This 
is the critical Ginzburg regime of large fluctuations. A non-equilibrium 
transition with a finite $\tau_Q$ results in an additional non-equilibrium 
density $dn(\beta)~=~n(\beta)-n_{eq}(\beta)~>~0$. KZM predicts that
\begin{equation}
dn_{\rm KZM}(\beta)~\approx~
\xi_Z^{-1}~=~
\tau_Q^{-\frac{\nu}{y+1}}~\approx~
\tau_Q^{-0.324\pm 0.003}.
\label{dnKZM}
\end{equation}
Before this prediction is compared with the numerical data in Fig.\ref{n}, let 
me digress on annihilation of domain walls.

\begin{figure}
\epsfig{file=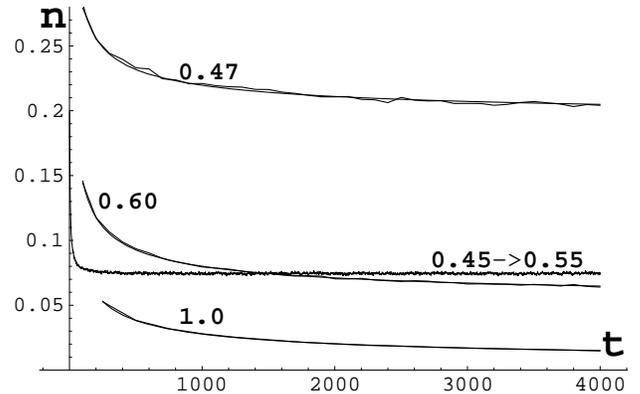, width=8.2cm, clip=}
\vspace*{0.1cm}
\caption{Density of defects $n(t)$ starting from an initial state 
with random spins and decaying towards $n_{eq}(\beta)$ for
$\beta=0.47,0.6,1.0$. The ''$0.45\to 0.55$" marks the plot of $n(t)$
starting from the state of equilibrium at $\beta=0.45>\beta_c$ in the 
Ginzburg regime and decaying quickly towards a new equilibrium at 
$\beta=0.55$.}
\label{ann}
\end{figure}

{\bf Defect annihilation.---} First example is annihilation of defects
from an initially totally random spin configuration. The initial $dn(t=0)$
decays in time. Fig.\ref{ann} shows the equilibrating $n(t)$ for several
values of $\beta>\beta_c$. Each decay is fitted with a solid line that
follows the power law $dn(t)=(\tau_a/t)^{1/2}$ with an exponent of $1/2$
known from the theory of phase ordering kinetics \cite{Bray}. The best
fits are $\tau_a=0.86\pm 0.05,0.93\pm 0.05,0.64\pm 0.05$ for
$\beta=0.47,0.60,1.0$ respectively. They are more or less constant in the
considered range of temperatures: as the critical point is approached the
time scale for annihilation $\tau_a$ neither diverges (as suggested in
Ref.\cite{Rivers}) nor vanishes (as suggested in Ref.\cite{Bett}), but
remains finite and close to the microscopic $\tau_0=1$,
\begin{equation}
\tau_a~\approx~\tau_0~.
\end{equation}  
The quench time $\tau_Q$ determines the time available for defect annihilation. 
At late times after the transition, when most of the initial KZM domain walls 
are already annihilated, we expect the scaling
\begin{equation}
dn(\beta)~\approx~
\left(\frac{\tau_0}{\tau_Q}\right)^{\frac{1}{2}}~.
\label{dna}
\end{equation}
It also follows from a phenomenological equation:
$\tau_0\frac{d}{dt}dn(t)=-\frac12 dn^3(t)$. Its solution is
\begin{equation}
dn(t)=\frac{dn(0)}{\sqrt{1+\frac{t}{\tau_0}dn^2(0)}}~.
\label{Vinen}
\end{equation}
Note that at late times $dn(t)$ is forgetting its initial value 
$dn(0)=dn_{\rm KZM}$. This solution is an ilustration of the exact
result (\ref{dna}) from Ref.\cite{Bray}.

Second example is annihilation of domain walls from an initial state of
equilibrium at $\beta>\beta_c$. The initial state was prepared by starting
from fully polarized spins, all $S_i=1$, and then heating them up at
$\beta=0.45$ for a time of $10^5$ sufficient to reach thermal
equilibrium with $n_{eq}(0.45)=0.20$. Then at $t=0$ $\beta$ was suddenly
increased (the heat bath was cooled) to $\beta=0.55$. Fig.\ref{ann} shows
$n(t)$ decaying towards the new equilibrium at $n_{eq}(0.55)=0.075$. This
decay is much faster than for random initial spins because the equilibrium
domain walls in the Ginzburg regime at $\beta>\beta_c$ are just boundaries
of bubbles of the minority spin-down phase in the spin-up
polarized ferromagnet. The bubbles together with their walls decay soon
after the temperature is turned down.

\begin{figure}
\epsfig{file=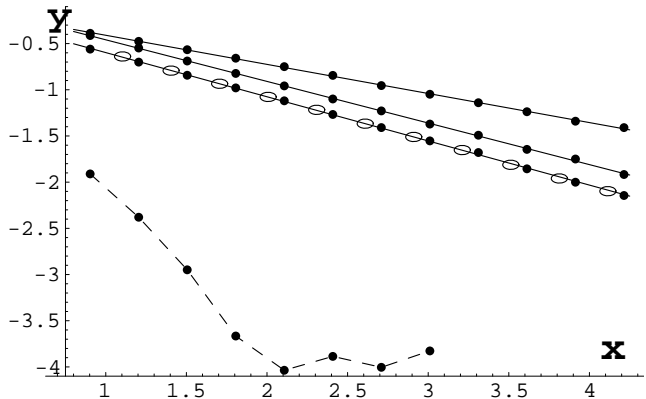, width=8.5cm, clip=}
\vspace*{0.1cm}
\caption{$y=\log_{10}dn(\beta)$ as a function of $x=\log_{10}\tau_Q$ for
$\beta=0.4407,~1.0,~1.5$ from top to bottom. Solid lines are the best
linear fits with slopes of $-0.315\pm 0.007,-0.45\pm 0.01,-0.48\pm 0.01$ 
respectively. Circles show
$dn(\beta=1.5)$ in a quench starting from $\beta_0=0.6$ and random initial 
spins. The points connected by a dashed line
show densities $dn(\beta=1.5)$ in a quench starting from
$\beta_0=0.47$ in the Ginzburg regime and spins initially in thermal 
equilibrium.}
\label{scaling}
\end{figure}

{\bf KZM versus annihilation.---} Figure \ref{scaling} is a double
logarithmic plot of the non-equilibrium density $dn(\beta)$ in Fig.\ref{n}
as a function of $\tau_Q$ for a number of $\beta$s. The slope at the
critical $\beta_c=0.4407$ is $-0.315\pm 0.007$. This slope is consistent
with the KZM slope (\ref{dnKZM}) of $-0.324$ and very different from a
mean-field KZM slope of $-0.65$ for $\nu_{\rm MF}=1/2$. The initial
non-equilibrium density of domain walls is determined by KZM.

In contrast, similar slopes for $\beta=1.0$ and $1.5$ are $-0.45\pm 0.01$
and $-0.48\pm 0.01$ respectively, and they are consistent with the phase
ordering kinetics exponent of $-1/2$ in Eq.(\ref{dna}). Apparently at
later times the system forgets the initial density $dn_{\rm KZM}$ and
$dn(\beta)$ is determined solely by the dynamics of defect annihilation.

Indeed, the circles in Fig.\ref{scaling} show $dn(\beta=1.5)$ for a
simulation where $\beta(t)$ is ramped up like in Eq.(\ref{ramp}), but
starting from the initial $\beta_0=0.6>\beta_c$. The spins were random at
the initial $\beta_0$. The circles sit on the solid line which is a fit to
$dn(\beta=1.5)$ obtained from a full quench like in Eq.(\ref{ramp}).  The
annihilation dominated $dn(\beta)$ at later times is not sensitive to the
details of the KZM of defect formation, compare
Eqs.(\ref{dna},\ref{Vinen}).

However, the defects that survive annihilation at later times are KZM
defects quenched in from the disordered phase. As we have already seen,
compare Fig.\ref{ann}, that annihilation of the Ginzburg domain walls is
much faster than annihilation of defects from the initially random spin
state. The latter state contains large domain walls while in the former
domain walls are boundaries of bubbles of a minority spin phase.
The points in Fig.(\ref{scaling}) connected by a dashed line show
$dn(\beta=1.5)$ after a quench starting from the equilibrium state at
$\beta_0=0.47$ in the Ginzburg regime.  These densities are orders of
magnitude lower than densities from the full quenches starting at
$\beta=0$: Ginzburg defects do not survive annihilation.

{\bf Conclusion.---} I presented first numerical test of the Kibble-Zurek
mechanism (KZM) in the Ginzburg regime of large thermal fluctuations. In
this regime both the Zurek length $\xi_Z$ and the microscopic length
$\xi_0$ are relevant length scales that determine the density of defects.
However, the density of non-equilibrium defects frozen into the ordered
phase by a quench from the disordered phase is determined by $\xi_Z$ only.
This initial density of defects is gradually annihilated and when the
system leaves the Ginzburg regime the density of defects is no longer
sensitive to the details of the KZM, but it is determined by the dynamics
of the annihilation process only. In particular, the dependence of the
density on the transition rate is determined by an exponent that comes
from the theory of phase ordering kinetics and not from the KZM. The only
way to see the KZM scaling (\ref{dnKZM}) directly is to measure the amount
of disorder close to the critical point where the non-equilibrium KZM
density is largely obscured by the prevailing equilibrium thermal
fluctuations. However, the defects that survive the annihilation are the
KZM defects quenched in from the high temperature phase, the defects
quenched in from the Ginzburg regime decay much faster. The surviving
defects are a clear, though indirect, signature of the KZM.


{\bf Acknowledgements.---} I appreciate discussions with Henryk Arod\'z,
J\'ozef Sznajd, and Wojciech \.Zurek. This research was supported in part
by the KBN grant 2 P03B 092 23 and in part by COSLAB.


\end{document}